# E-Courseware Design and Implementation Issues and Strategies

Shakeel Ahmad[1, 2], Adli Mustafa[2], Zahid Awan[3], Bashir Ahmad[1], Najeebullah[4] and Arjamand Bano[5]

[1]Institute of Computing and Information Technology Gomal University, Pakistan
[2]School of Mathematical Sciences, Universiti Sains Malaysia Penang, Malaysia
[3]Department of Business Administration Gomal University, Pakistan
[4]Department of Public Administration Gomal University, Pakistan
[5]Mathematics Department Gomal University, Pakistan

**Abstract -** Over the last few years electronic learning has been in use mostly by corporate institutes in the form of computer aided instructions and computer based training. The scope of such use has not only been limited to introductory courses for beginners and working people but also to impart knowledge in higher education sector. Due to increasing market demands and current prevailing law and order situation of this area (during which the University remain closed for uncertain period of time on many occasions) Gomal University D.I.Khan, Pakistan is planning to introduce e-learning at undergraduate and post graduate level in computer and management sciences for smooth and uninterrupted delivery of quality education to local and distant students. Obvious result of e-learning will be two fold. First it will meet market demands along with smooth uninterrupted delivery of quality education and secondly will solve the growing problem of shortage of experts raised by the current law and order situation. This paper investigates the main issues involved in designing and implementing an effective electronic courseware for students with diverse backgrounds belonging to this remote area. Some effective strategies for electronic delivery of courses to local and distant students are also presented along with some examples of implementation.

## 1. Introduction

The Institute of Computing & Information Technology (ICIT) and Department of Business Administration (DBA) at Gomal University Dera Ismail Khan (DIKhan) are recognised as main contributors of IT and business education in this remote area of North West Frontier Province (N-W.F.P) of Pakistan, which offers a variety of courses, both at undergraduate and postgraduate level, in order to meet a wide range market demands. Like other computing and business administration departments of the remote areas Universities in the country, these departments are also facing the problem of recruiting and retaining competent staff and because of current prevailing law and order situation of this remote area most of the time throughout the academic session University remain closed for uncertain period of time resulting wastage of students time and delay in academic session. While resolving these problems, the departments want to keep delivering smooth and uninterrupted quality education to students in order to meet the overall country demand.

Both departments have their own well established local intranet equipped with state of the art technology where all the teaching material for all modules will be available for the students. Apart from network infrastructure there is a need for excellent electronic courseware that not only meets the student's demands, but could also be used for the delivery of a particular module in the absence of a competent instructor. The technology and expertise to design such courseware is available. The main reason for the unsatisfactory current situation is the high cost of producing high quality electronic courseware and the missing opportunity to share this cost in a large user community [1]. Another reason for the unsatisfactory situation is the low quality of widely available courseware unable to fulfil the needs of students having diverse background. Once excellent electronic courseware is developed, it can then be used for distance learning in order to provide quality services as well as to generate external revenue for the University.

This paper discusses several issues regarding the need for, and preparation of, electronic courseware. It also presents some effective strategies for the electronic delivery of modules to local and distant students especially by keeping in mind the diverse background of the students of this remote area. Section 2 discusses the need for electronic learning. A brief presentation of certain strategies for the preparation of electronic courseware and delivery of courses follow in Section 3. Section 4 presents key people involved in preparing courseware and the users of this courseware. Section 5 presents few examples for the possible implementation of electronic learning in the ICIT and DBA. Finally Section 6 presents the conclusions.

## 2. Why E-Learning?

Electronic learning has been in use for several years in the form of computer-aided instruction (CAI) and computer-based training (CBT). These systems are commonly found in primary and secondary schools to teach children language, mathematics and typing skills while many corporate institutes use them as major means for the delivery of courses they offer. They have shown the ability to replace teaching for routine coaching tasks such as courses offered by the Virtual University of Pakistan [11], Allama Iqbal Open University, Pakistan [12], Cisco [13] etc. and have been found very successful. Intelligent tutoring (IT) is another approach that is being used for electronic learning, and this approach can be effectively used to improve the quality of learning for the electronic courseware. A considerable number of tools are now available to create interactive and effective courseware for electronic learning.

One of the main reasons for the use of electronic learning in ICIT and DBA is to provide a means for the smooth delivery of modules whose experts are not available and where only a minor help in terms of staff support is available. It will also help those students whom cannot attend class sessions due to prevailing law and order situation and whom want to catch-up in their studies. Live delivery of an interesting seminar by some outside speaker can be made possible or such seminars can be available on the web for the local and distant students.

Other main reasons include [1, 2, 3]:

(i) Faculty is the most expensive element in any educational programme. A well-established electronic courseware and the selection of appropriate technology can reduce the cost by increasing the student-staff ratio.
(ii) Due to the prevailing law and order situation, it's become difficult for both departments to recruit and retain the experts.
(iii) For smooth uninterruptible delivery of quality education to save the students time.
(iv) As University wants to generate external revenue, electronic learning can play a vital role in establishing franchised courses and provide the quality distance learning service.
(v) In case of distance learning, it can be used to generate a tremendous amount of income. In this way students from different social, cultural, economical and experiential background can be connected.
(vi) To run distance learning programmes, department will necessarily need additional staff. Once a richer and effective electronic courseware is established, will result a tremendous reduction in the cost, as it will displace most of the additional staff. The Virtual University of Pakistan is the best example to provide an evidence. All they need is the part-time tutors to provide some tutorial support to distant students. This tutorial support can be in the form of feedback on written coursework, assessment, response to students' queries, etc. A summary of several cases presenting evidence for reduction in cost can be found in [10].
(vii) Departments can play an important role in networking with other Universities having similar programmes and can share the resources, which will help in reducing cost for its preparation.
(viii) Short courses can be designed for the working people who want to get familiar with the new technology and cannot do the full time or part time degree courses.

## 3. How to build E-Courseware?

This section will present some strategies for developing an electronic courseware and also the people who are needed for this purpose.

### 3.1 How electronic learning is different from classroom teaching?

A classroom teacher observes the students whether they are attentive or not and adjust the course delivery, e.g., by re-planing class activities to involve students at greater degree, in order to meet the needs of a lecture for a particular teaching session. In contrast, for electronic learning, a teacher cannot have these observations directly. It is also very difficult to carry on a class discussion. Furthermore, if electronic learning is being used for distance learning, a teacher may not know the different educational and social backgrounds of a diverse range of students.

**Students' needs**

In order to meet students' needs and to motivate them, efforts must be made in terms of instructional contents and effective learning styles. Following steps can be taken for this purpose [3]:

- helping students to get familiar with the technology used for delivery,
- making arrangements to respond promptly to the students' queries regarding contents,
- be aware of the students' needs in meeting standard University deadlines,

- making sure that students take an active role in electronic course delivery by independently taking responsibility for their learning,
- instructor should interact with individual students,
- instructor can distribute questionnaires and based on their feedback and interaction, he/she can meet the individual needs,
- Students should be given detailed remarks on their coursework and should be referred to additional resources for supplementary information.

**Effective teaching skills**
Whilst planning for the teaching skills to be used for electronic learning, main focus should be students rather than delivery technology. Instructors must be aware of the students learning styles because some will learn quickly than others. Instructors should use short statements in instructional contents and should ask direct questions. Students should be encouraged to use emails very frequently and with confidence to sort out their problems regarding understanding of instructions. They should also be encouraged to use software tools available for on-line chat in order to establish some kind of group discussions.

**Preparation of instructions**
Once the course requirements and students' needs have been carefully studied, instructors' task is to select among all the technologies available for preparing effective instructions. It involves an analysis to see what technologies and tools are available in terms of learners' needs and course requirements (for more detail see [6, 7, 9]). There is a great need for someone to collect and organise resources that can be used by the instructor to develop instructions. Links to relevant useful web resources should also be collected for additional help to give students better understanding. This collection of resources should be constantly updated.

**Courseware Management Tools**
A wide range of tools is available to create electronic courseware [4, 9]. Particularly, the multimedia courseware authoring system vendors provide a sufficiently wide range to cover the main categories of products required for web-based training, such as WWW publishing, internet-based video conferencing, integrated distributed learning environments, etc. as identified in [6]. The most difficult part in preparing electronic courseware is the selection of appropriate tools. The selection process requires an in-depth analysis of the programme requirements and strengths of the tools available.

Comprehensive list of the elements that have to be considered for this purpose can however be found in [6].

**3.2 Strategies**
Following are some of the strategies that have been proposed in the literature [3, 5] and can be used to develop instructions for electronic learning:

**Recorded Lectures**
A very simple way of teaching electronically is using recorded lectures. These can be live or pre-recorded. A little effort involves in recording a lecture. For delivery, all it needs is either video streaming or slide presentation with synchronised audio clips. These recorded lectures cannot be effective, as they do not have the strength of interactive teaching. Still this approach can be very useful for delivering seminar by outsiders or putting an interesting seminar on the web for the out world.

**Multi-Media software**
Another approach is to use multi-media tools available to create interactive movies, such as Adobe Director (formerly Macromedia Director), Flip Camcorder etc. The movies developed by using these tools can then be played in a Web browser. Although such interactive movies can be most useful way of preparing and delivering electronic courseware, they involve a lot of cost for their production in terms of tools and highly trained staff.

**Computer-Based Training**
A considerable number of tools are available to create instructions for computer-based training programmes. These include Microsoft PowerPoint, Adobe Authorware, Dreamweaver, Flash, SumTotal, Toolbox, BlackBoard etc. These types of tools are more popular in preparing tutorials and self-contained educational modules.

**Web Page Development Tools**
Most of the tools, which previously produced CBT courses, can now be used to produce instructions for Web delivery. Instructions can be created using Web's technologies like HTML, XML, CGI, Java Script, PHP etc. Web based courses are generally more organised as they can help learners to focus their studies by using hyperlinks and to integrate material available on different sites.

**Collaboration Tools**
One of the main requirements of electronic learning is to provide a means of collaboration between learner and facilitator. For a real time environment, this collaboration is supposed to be synchronised where learner and facilitator must be online at the same time. In time delayed environments, where learner and facilitator logged on at different times, this collaboration is known as asynchronous. A number of tools are available to support these collaborations such as Microsoft NetMeeting, Skype, Twitter, Voice Thread etc. for synchronous collaboration and Lotus Notes, Wizlite, Noteclip etc. for asynchronous collaboration. More details about collaboration tools can be seen in [14].

**Web-based training**
Despite of all the expected improvements in communications to make web-based training more efficient and cheap, still there are certain disadvantages in web-based training [4]. The main disadvantage is the overhead involved in acquaintance with new tools, each time, adopted for course management. In spite of all disadvantages, there are several advantages in using web as source for course delivery [2 ,4]:
- A trainee programme can be designed with little dependence on the number of trainees, which is of a particular importance when this number is very large or difficult to estimate over a large period of time
- Trainees are able to make the best use of their time, according to their occupations at any time and to need they feel to improve their professional skills.

Many institutes including Information and Educational Technology (IET), University of California [5] and Engineering Outreach, University of Idaho [8] use web pages to create and support an information technology environment that enhance the ability of their communities to teach, do research and provide public services.

## 4. Users of the E-courseware and their roles
Successful electronic learning programme relies on the integrated efforts of students, faculty, support staff and administrators. Following is the brief description of their roles:

**Students**
The most important part of the electronic learning is to meet students' needs. Efforts for the success of electronic learning can be measured on the basis of some tests to see how students are receiving instructions. The primary role of a student is to learn which requires planning, motivation and ability to analyse. He/She is also responsible for applying the knowledge he/she gained through electronic learning.

**Teachers**
Teachers play the key role in making the electronic learning successful. For a traditional interactive lecture they plan the contents to be delivered, make the objectives clear and prepare some class activities to judge the understanding of students as well as to involve them. They are ready to help if some question arises. Special challenges confront teachers preparing instructions for electronic delivery. Such tasks involve:
- preparing instructional contents by taking into consideration the needs and expectations of students with diverse backgrounds
- developing a working understanding of delivery technology
- acting as a skilled facilitator and content provider

**Administrators**
They are the most influential people involved in planning and decision making. They have to make sure that the electronic learning is achieving its objectives while keeping the required standards. They work very closely with academics and support staff to ensure the effectiveness of instructional contents and technological options selected for the success of electronic learning.

**Support Staff**
They play very important role in collecting and organising resources for the efficient preparation and smooth delivery of instructional contents. Without skilled support staff success of electronic learning is very difficult.

## 5. Implementation
Although its difficult to say anything about the success of electronic learning for full-time or part-time students in the University, literature [5, 8, 9, 11] gives enough evidence of its success for various distance learning programmes. Following are some examples presenting implementation of electronic learning in the ICIT and DBA Gomal University DIKhan:

**Web Page Design and Development**
This module is usually offered in semester one to first level students of different programs offered by ICIT and DBA. The objective of this module is to introduce several software tools that students will use in the design and development of web page. This module does not need any kind of lecture sessions. The only important element in its delivery is the design of effective laboratory tutorials and exercises for assessment. Once these have been designed, a little lab assistance will make its efficient delivery possible.

### Object Oriented Modelling and Design/ Introduction to Business Finance

These modules can also be delivered using electronic courseware. These are very technical subjects and the module co-ordinators have to make sure that their electronic delivery is successful in terms of students understanding. For this reason, very careful exercises need to be designed and a student should not be allowed to proceed until he/she has met the minimum criteria. The preparation of electronic courseware for these modules needs much more expertise than the previous example.

Similarly, electronic courseware for almost every module can be prepared and could be used with very little staff involvement. The only problem is how to involve students? Experience shows that students use Internet mostly for entertainment and very little for learning purpose. Interesting examples, exciting exercises and impressive presentation of electronic material may help to solve this problem to some extent. If, somehow, this problem is resolved, electronic learning can become very effective in ICIT & DBA and can substantially help in reducing staff teaching load. Once electronic courseware is developed, these departments will be able to provide quality service to distance students whilst minimizing chances of possible complaints regarding availability of online support. University can start distance learning courses to generate revenue. In addition after the successful implementation of distance learning this service may further be extended to different areas of study.

### 6. Conclusion

This paper presents some of the major issues regarding importance and establishment of an electronic learning environment in two main departments of the Gomal University Pakistan. These issues include the need for such environment, the efforts required for its development, the strategies for its implementation and finally the roles of various people involved in preparation and its use. The major reasons for the electronic learning environment are:
a) To improve the resources available for the students
b) Possibly to offer a module for which an expert staff is not available.
c) To generate external revenue under distance learning programmes and finally
d) For smooth uninterruptible delivery of quality education.

Establishing such an environment needs a great deal of analysis of the requirements and the technologies. If the electronic learning environment does not reduce a significant amount of cost, which is otherwise needed for the faculty, then it is of less interest to the University. The technology and strategies must be chosen very carefully so that they not only return with quality of resources but also with a great amount of reduction in the cost.